\documentclass[aps, pra, reprint, longbibliography, superscriptaddress]{revtex4-1}

\usepackage{hyperref}
\usepackage{graphicx}
\usepackage[utf8]{inputenc}
\usepackage{xcolor}
\usepackage{amsmath, amssymb}

\begin{document}

\title{Reply to the Comment on ``Berezinskii–Kosterlitz–Thouless Transition in 
Two-Dimensional Dipolar Stripes" }

\author{Raúl Bombín}
\email{raul.bombin@ehu.eus}
\affiliation{Departament de F\'{i}sica, Universitat Polit\`{e}cnica de Catalunya, Campus Nord B4-B5, E-08034, Barcelona, Spain}
\affiliation{Departamento de Física de Materiales, Facultad de Químicas, UPV/EHU, Apartado 1072, 20080 San Sebastián, Spain}
\affiliation{Centro de Física de Materiales CFM/MPC (CSIC-UPV/EHU), Paseo Manuel de Lardizabal 5, 20018 Donostia-San Sebastián, Spain}

\author{Ferran Mazzanti}
\email{ferran.mazzanti@upc.edu}
\affiliation{Departament de F\'{i}sica, Universitat Polit\`{e}cnica de Catalunya, Campus Nord B4-B5, E-08034, Barcelona, Spain}

\author{Jordi Boronat}
\email{jordi.boronat@upc.edu}
\affiliation{Departament de F\'{i}sica, Universitat Polit\`{e}cnica de Catalunya, Campus Nord B4-B5, E-08034, Barcelona, Spain}

\begin{abstract}
This is a  Reply to the  Comment from F. Cinti and M. Boninsegni on our recent work on the Berezinskii-Kosterlitz-Thouless (BKT) phase transition in a two-dimensional dipolar system [R. Bombín, F. Mazzanti and J. Boronat, Physical Review A \textbf{100}, 063614 (2019)]. The main criticism about our work, expressed in that Comment, is that we did not explicitly report the two spatial contributions to the total superfluid fraction. Here, we analyze our results for a point of the phase diagram corresponding to the stripe phase, close to the gas to stripe transition line, and for a temperature below the BKT critical temperature.  The scaling with the system size of the contribution to the superfluid fraction, coming from the direction in which spatial order appears, shows that it remains finite in the thermodynamic limit, as we already stated in our original work. This allow us to state that the stripe phase is superfluid at low temperatures. Furthermore, we offer some comments that help to understand where the differences between the results of Cinti and Boninsegni and ours come from.
\end{abstract}

\maketitle

In the Comment by Cinti and Boninsegni~\cite{cinti2020comment} (in the following refereed as  “the Comment” for the sake of simplicity), their authors criticize our recent work \cite{Bombin2019} on the analysis of a BKT transition  in a two-dimensional (2D) dipolar system. The arguments provided there constitute an extension of those described in Ref~\cite{Cinti2019},  where one of our previous works~\cite{Bombin2017}, regarding the study of the same system but at zero temperature, was also criticized. Indeed, no new results are presented in the recent Comment.

The authors of the comment already presented their finite-temperature calculations~\cite{Cinti2019} to confront them with our zero-temperature 
ones~\cite{Bombin2017}. More recently, we have extended our research on this system to finite temperatures \cite{Bombin2019}, finding very good agreement with similar works \cite{Filinov2010} and with our previous study in the zero-temperature limit. Although both references \cite{Bombin2019} and \cite{Cinti2019} use the Path Integral Monte Carlo (PIMC) technique to obtain finite-temperature predictions, some remarkable technical and methodological differences can explain the different conclusions reached in each case. We start commenting them.

First, from the technical point of view, the authors of Ref.~\cite{Cinti2019} explicitly state that they use the simplest action in their calculation, that is the Primitive one. This action is only accurate up to order $(dt)^2$, with $dt$ the (imaginary) time-step of the simulation. On the other hand, and as stated in our work, we use one of the $(dt)^4$ actions of Ref.\cite{Takahashi1984,Takahashi1984a,Chin2002,Chin2004,Sakkos2009}. In fact, we have tried several of them to finally adopt the one that shows the smallest variance in the results. This is an important issue because a better action allows using a considerable reduced number of beads (intermediate integration coordinates) in the simulation,thus allowing to obtain reliable results at the very low temperatures that are of interest. Even worse, the Primitive action is known to suffer from critical slowing down, which can lead to biased results and make the simulation impractical, or totally unfeasible, in some cases. Such a simple action implies that the number of required beads (intermediate coordinates in the simulation chain) raises significantly, and sampling them ergodically along a finite-time simulation becomes a formidable challenge. On the contrary, a fourth-order action requires much less beads and the error in the estimation converges much more rapidly. All in all, at least in the case of quantum Monte Carlo simulations, the Primitive action has been largely superseded by other, more accurate actions like the ones we currently use.
As a final remark on this point, some of the temperatures used in Ref \cite{Cinti2019} are as low as $T/nr_0^2\sim$0.08$\varepsilon_0$ with $n$ the density of the system, $T$ the temperature  and with  $r_0$ and $\varepsilon_0$ the dipolar units of length and energy (see, for example, Ref.~\cite{Bombin2019} for their definition). In our work, we find that the critical temperature $T_{BKT}$ for the stripe phase is around $T/nr_0^2=$ 0.6 $\varepsilon_0$, and we had to use more than 150 short-time intermediate propagators (accounting for more than 450 beads) in order to properly converge with the improved fourth order action that we use. Taking into account that the number of required short-time propagators scales with the inverse of the temperature, achieving reliable results for temperatures as low as the one quoted above would require in our case the use of no less than 1000 propagators. And the situation would be dramatically  worse if one employs the Primitive action, as is done by the authors of Ref \cite{Cinti2019}. In our work, calculations are restricted to temperatures $T/nr_0^2>$0.35 $\varepsilon_0$.

Second, it is remarkable that in Ref \cite{Cinti2019}, the authors performed “a few targeted” simulations at some points and temperatures across the phase diagram, while in our work we have systematically studied the system at different temperatures and system sizes. By employing the scaling laws of the Berezinskii-Kosterlitz-Thouless (BKT) phase transition, we determine the critical temperature at which the transition from normal fluid to superfluid occurs in the stripe phase. In this sense, our exploration is much more exhaustive and we systematically recover consistent results along the whole phase diagram. Moreover, the method is the same as the one employed for the study of the isotropic gas phase of the same system in Ref.~\cite{Filinov2010}, whose results we are able to recover (See Ref.~\citep{Bombin2019}).

In the following, we reply to the specific comments raised about our work.
\begin{enumerate}
\item  We state that for points in the phase diagram close to the gas-stripe transition line, the superfluid signal $\frac{\rho_s}{\rho}$ is large , the major contribution being the one coming from the stripe direction $\frac{\rho_s^X}{\rho}$ but with finite values also in the transverse one $\frac{\rho_s^Y}{\rho}$.  We defined these two contributions as follows
\begin{equation}
\frac{\rho_s}{\rho}=\frac{1}{2}\left[\left(\frac{\rho_s^X}{\rho}\right)+\left(\frac{\rho_s^Y}{\rho}\right)\right].
\end{equation}
following Ref.~\cite{Bombin2017}, where the two contributions to the superfluid fraction where already reported in the limit of zero temperature. The authors of the Comment state that our results of Fig. 4 of Ref \cite{Bombin2019} contradict the statement of having a finite $\frac{\rho_s^Y}{\rho}$ contribution to the total superfluid fraction, as the largest values of $\frac{\rho_s}{\rho}$ are close to 0.5. However, our figure exhibits a clear increasing tendency as the temperature is decreased, and clearly the only meaningful  extrapolation would yield values larger than 0.5. Moreover, having $\frac{\rho_s^X}{\rho}= 1$ for temperatures close to the BKT transition temperature $T_{BKT}$ as the authors of the Comment suggest, would be quite unrealistic. To discard further concerns regarding this point, in Fig.~\ref{fig:scaling_N} we show the scaling with the system size of the superfluid density, measured across the transverse direction $\frac{\rho_s^Y}{\rho}$, and evaluated with different number of particles $N$, keeping the density nr$_0^2$ = 128 and polarization angle $\alpha = 0.6$ fixed.

\begin{figure}
\includegraphics[width=1\linewidth]{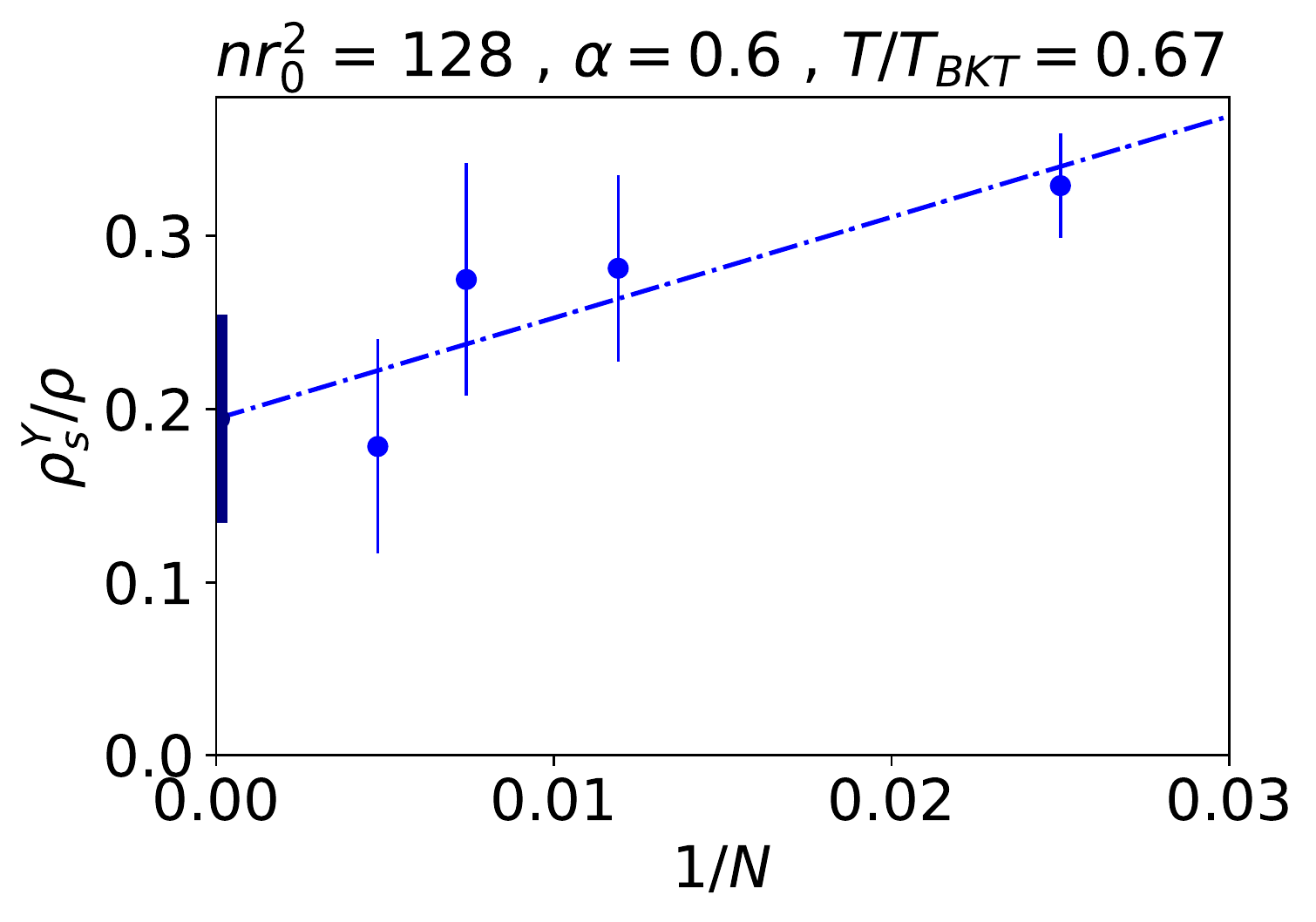}
\label{fig:scaling_N}
\caption{Finite-size scaling of the superfluid density $\frac{\rho_s^Y}{\rho}$, for a point of the phase diagram in which the stripe phase is present and close to the gas-stripe transition line (nr$_0^2$ = 128 and $\alpha = 0.6$). An estimation of the extrapolated value, obtained with a linear fit to the data (dashed line) is shown on the vertical axis. }
\end{figure}

Moreover, for the density $nr_0^2=$256 and polarization angle $\alpha =$ 0.6 (a point in the phase diagram that is far away from the gas-stripe transition line, lying in the deep stripe regime), we have checked that almost perfect agreement exists between the finite temperature PIMC results and the zero temperature  ones of  Ref \cite{Bombin2017}. Quite remarkably, our calculations in Ref. \cite{Bombin2017} predict a zero-temperature superfluid fraction of 0.54(5), which is in perfect agreement with the low temperature prediction shown in Fig. 3 of Ref. \cite{Bombin2019}. This is a strong benchmarking test for both the methods and for our codes, as different techniques and formalisms are used in each case.  We do not find any reason to believe that this agreement could not be extended to the rest of points of the phase diagram.

\item In the Comment, it is argued that it is not possible "to provide reliable numerical predictions for systems comprising just a few particles". This assertion is surprisingly wrong, as finite-size scaling relations have been widely employed in all fields of physics to describe (quite successfully) the properties of  physical systems. Moreover, the results that the authors of the Comment presented in Ref.~\cite{Cinti2019}, are done with a number of particles that is similar to ours. In any case, there are precise scaling laws that allow performing extrapolations to the thermodynamic limit, previously reported and successfully employed, for instance in Ref.~\cite{Filinov2010}. Remarkably, our results are in agreement the ones in this same reference, in the zero polarization angle limit (isotropic case).

\item In the Comment, it is also stated that the one-body density matrix (OBDM) can be very different along the two perpendicular directions of the plane, and thus biased by its larger value if one evaluates a circularly averaged estimation. In our previous work, we showed that when the OBDM is expanded into partial waves only the zero mode (corresponding to the circularly averaged one-body density matrix) contributes at long distances, and that for ($N>$100) the values of the  $X$ and $Y$ component are compatible at $r=L/2$, with $L$ the length of the simulation box. In much the same way, we can not understand that the OBDM's reported by the authors of the Comment in Ref.~\cite{Cinti2019} do not show any noticeable dependence on the temperature,  even when they vary its magnitude by a factor of four. Related to this, another difference between the two works is appreciated when one looks at the snapshots from the simulations. In our case, these images of the system show interchanges between the different stripe lines, at odds to what one can appreciate having a look at the ones reported in Ref.~\cite{Cinti2019}. 


\item Finally, the Comment is closed with a strong assertion, for which no arguments are provided: ``More generally, we reiterate here our contention that no supersolid phase of dipolar bosons exists in 2D, the third dimension being required for the stabilization of such a phase". We would like just  to point out that supersolid phases have been reported in other two-dimensional systems, for example for the 2D dipolar system on the lattice on Ref.~\cite{Bandyopadhyay2019} and for a system with finite-range interactions  also in a 2D lattice in Ref.~\cite{Masella2019}. In the later case, calculations where performed employing the PIMC method, as in our work.

\end{enumerate}

In summary, we have shown that for temperatures $T<T_{BKT}$ the contribution to the superfluid fraction coming from the transverse direction is finite, even in the thermodynamic limit. Therefore, the conclusions that we reported in Ref \cite{Bombin2019}, in particular the ones regarding the supersolid character of the stripe phase, are fully valid. Furthermore, our results regarding the superfluid fraction, the one-body density matrix and the interchanges between different stripes are in severe disagreement with those presented by the authors of the Comment in Ref.~\cite{Cinti2019}. We strongly believe that the results presented there are biased by the low-order action that they use.

\acknowledgments{
This work has been supported by the Ministerio de
Economia, Industria y Competitividad (MINECO, Spain) under grant No. 
FIS2017-84114-C2-1-P.
 }


\bibliography{refs}

\end{document}